# Drift-free characterization of electro-optic tuning efficiency in lithium niobate photonic nanocavities

*Erqi Zhang,[1] Danyang Yao,[1*] Xu Ran,[1] Yiwei Zhang,[1] Duomao Li,[1] Youbin Wang,[1] Zhixuan Hu,[1] Jiaren Song,[1] Xiaoli Lu,[1] Xiaohua Ma,[1] and Yue Hao[1]*

[1]State Key Laboratory of Wide-Bandgap Semiconductor Devices and Integrated Technology, Faculty of Integrated Circuits, Xidian University, Xi'an, 710071, China;

* dyyao@xidian.edu.cn;

**ABSTRACT:**

Lithium niobate photonic crystal nanobeam cavity (PCNBC) represents a premier platform for integrated electro-optics, offering deep sub-wavelength mode confinement, enhanced light-matter interactions, and ultralow power consumption. However, accurate characterization of the electro-optic (EO) tuning efficiency in such high-Q devices is fundamentally impeded by DC drift, a time-dependent spectral instability arising from charge redistribution, surface screening, or buffer layer relaxation under sustained electric fields. Here, we report the systematic analysis of DC drift dynamics in lithium niobate nanocavities and demonstrate that conventional quasi-static DC voltage scanning yields highly unreliable characterization data. To circumvent this limitation, we introduce a drift-free, dynamic measurement methodology that employs high-frequency triangular-wave voltage sweeps to effectively decouple the instantaneous electronic Pockels response from slow charge-relaxation processes. Validated across 35 devices with varying electrode geometries, our method delivers reproducible tuning efficiency of 4.3-4.5 pm/V with a low coefficient of variation of 1.1%, showing excellent quantitative agreement with three-dimensional finite-element simulations. This robust, drift-free measurement technique establishes a rigorous standard for the characterization and optimization of resonant cavity electro-optics, accelerating the development of high-performance thin-film lithium niobate photonic integrated circuits.

## ■ INTRODUCTION

Lithium niobate (LN) stands as a premier material platform for integrated photonics, distinguished by its exceptional electro-optic (EO) properties, broadband transparency, and compatibility with advanced nanofabrication[1]. Its large Pockels coefficient enables efficient voltage-controlled refractive index modulation, positioning LN as the material of choice for high-performance EO modulators spanning applications from telecommunications to microwave photonics[2]. The advent of lithium-niobate-on-insulator (LNOI) has revolutionized the field, enabling compact, low-loss photonic integrated circuits that define the frontier of next-generation photonic devices[3].

Cavity-enhanced EO modulators represent a particularly promising direction, exploiting the pronounced light-matter interaction within high-Q optical resonators[4,5]. Microring resonator (MRR)[6] and photonic crystal nanobeam cavity (PCNBC)[7,8] confine optical energy in sub-wavelength volumes while extending photon lifetimes, thereby amplifying the effective interaction strength between light and the EO medium. This enhancement yields reduced device footprints, diminished switching voltages, and lower power consumption, essential attributes for scalable photonic integration and energy-efficient optical interconnects[9-12]. The ultra-narrow resonance features intrinsic to high-Q cavities provide exceptional sensitivity to refractive index perturbations, making them compelling platforms for precision tuning and modulation[13].

A fundamental requirement for the development and optimization of these devices is the accurate characterization of the EO tuning efficiency η[14], defined as the resonance wavelength shift per unit of applied voltage, is essential for calculating critical performance indicators, such as the voltage required for π phase shift[15-18], the voltage-length product and the 3dB modulation voltage[19]. However, accurate characterization of η in LN photonics remains a significant challenge due to the inherent DC drift effect[20-22]. This phenomenon, typically attributed to surface charge screening or buffer layer instabilities, causes a time-dependent shift in the transmission spectrum under an applied electric field[23]. In high-Q nanocavities, where the transmission resonance linewidth is remarkably narrow, even minor drift-induced spectral instabilities introduce severe tracking errors during characterization[24,25]. As a result, conventional static or quasi-static voltage-scanning methods often yield unreliable data[26], yielding highly inaccurate and non-reproducible measurements.

In this work, we present a robust, drift-free characterization methodology designed to accurately quantify the tuning efficiency η of high-Q LN photonic nanocavities. Our approach employs a polymer-loaded PCNBC fabricated on the LNOI platform[27], combining the strong field confinement of photonic crystal structures with the high Q-factor necessary for sensitive EO measurements. Crucially, we implement a dynamic measurement technique based on triangular-wave voltage sweeps, which effectively mitigates the DC drift effect by continuously modulating the applied voltage. This dynamic technique allows for the extraction of the instantaneous voltage-dependent resonance shift, separating the true EO response from drift-induced errors and producing an accurate, drift-compensated measurement of the tuning efficiency η. The proposed method thus provides a robust pathway for reliable characterization of LN-based cavity devices, facilitating more accurate design and performance prediction of next-generation EO modulators.

## ■ PRINCIPLE OF DRIFT-FREE ELECTRO-OPTIC CHARACTERIZATION

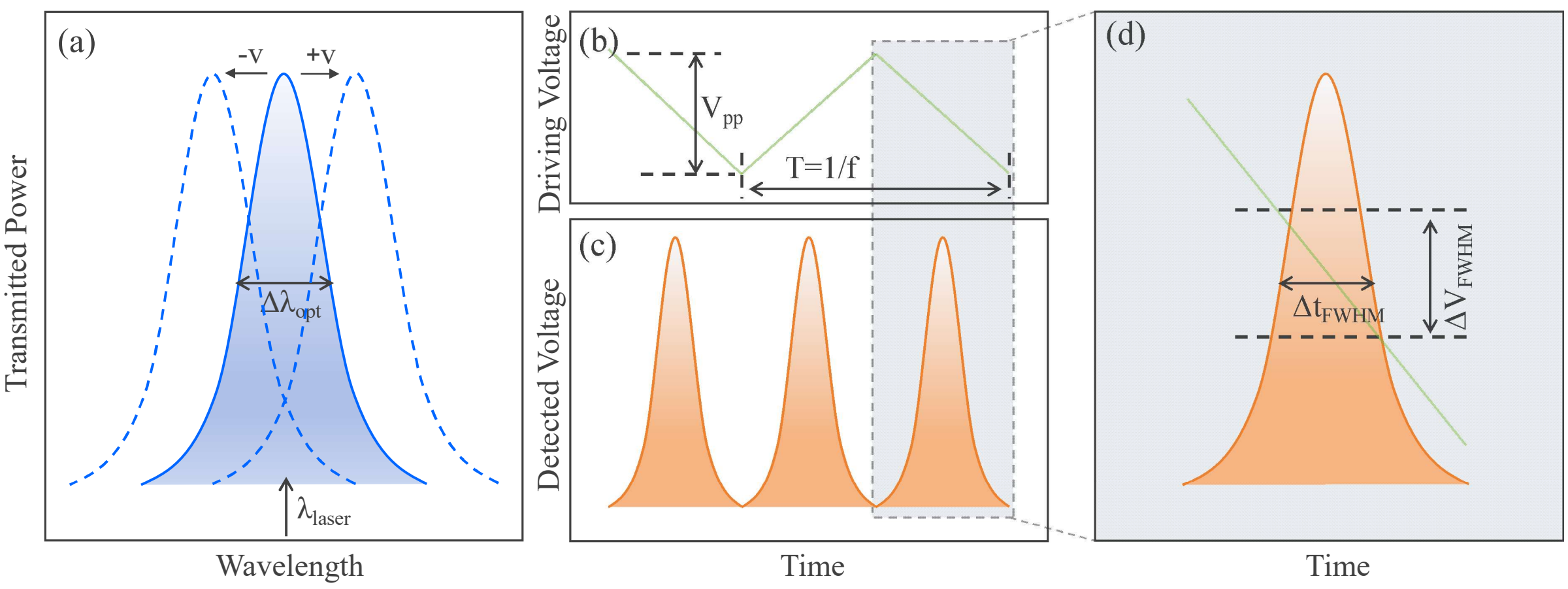

Figure 1. (a) Schematic of the cavity transmission spectrum under zero and non-zero bias. (b) Time-domain profile of the driving triangular-wave voltage. (c) Resulting time-domain optical transmission pulses detected by the photodetector as the cavity resonance sweeps across the fixed laser wavelength. (d) Magnified view of a single time-domain transmission pulse with a Lorentzian profile.

Conventional characterization of η in LN resonant devices relies on quasi-static DC voltage sweeps, where the resonance wavelength shift is tracked as a function of applied bias. In this approach, a series of static voltages are applied sequentially, and at each voltage setpoint the optical transmission spectrum is acquired by scanning the laser wavelength across the cavity resonance. The tuning efficiency η is then extracted from the slope of the measured resonance wavelength versus the applied voltage. However, this approach encounters fundamental, often unaddressed, limitations when applied to high-Q LN PCNBC under static electric fields. Under a static or slowly varying electric field, the instantaneous resonance wavelength $\lambda_{res}(t)$ is no longer uniquely determined by the Pockels effect, but follows a multi-component temporal evolution,

$$\lambda_{res}(t)=\lambda_0+\eta V(t)+\lambda_{drift}(t) \tag{1}$$

where $\lambda_0$ is the unperturbed resonance wavelength at zero bias, η is the intrinsic EO tuning efficiency η to be calibrated, V(t) is the applied voltage, and $\lambda_{drift}(t)$ is the parasitic, time-dependent drift term arising from charge redistribution, surface screening, or buffer layer relaxation under the sustained electric field. This drift term is entirely independent of the instantaneous applied voltage and evolves on its own characteristic relaxation timescale.

In high-Q nanocavities, the optical linewidth $\Delta\lambda_{opt}=\lambda_0/Q$ is remarkably narrow, typically on the order of a few picometers. Consequently, even a minor drift displacement spanning a fraction of the linewidth introduces a spectral shift comparable to or larger than the intrinsic EO response. If the dwell time at each voltage step or the interval between consecutive measurements is comparable to or longer than the drift relaxation time, the recorded $\lambda_{res}(t)$ conflates the true EO response $\eta\Delta V$ with the accumulated drift $\lambda_{drift}(t)$, leading to systematic overestimation or underestimation of $\eta$. The longer the voltage dwell time, the larger the drift-induced error, making the conventional DC scanning method fundamentally unreliable for high-Q LN nanocavity characterization.

The key physical insight enabling our drift-free characterization is the vast separation of timescales between the intrinsic Pockels effect and the DC drift relaxation. The Pockels (linear EO) effect is an electronic nonlinearity mediated by bound electron polarization, with a response time on the order of femtoseconds. This renders it effectively instantaneous across all practically relevant modulation frequencies. In contrast, the DC drift phenomenon, regardless of its microscopic origin, exhibits characteristic relaxation timescales ranging from milliseconds to seconds at room temperature.

Our methodology exploits this fundamental difference in response timescales by employing a triangular-wave voltage drive at frequencies far exceeding the characteristic frequency of drift. A narrow-linewidth continuous-wave (CW) laser is fixed at the cavity resonance wavelength $\lambda_{laser}=\lambda_0$ under zero bias (Figure 1a). When the triangular voltage V(t) is applied to the modulator electrodes, the cavity resonance sweeps linearly in wavelength as,

$$\lambda_{res}(t)=\lambda_0+\frac{\eta V_{pp}F_{tri}(t)}{2} \quad (2)$$

where $V_{pp}$ is the peak-to-peak amplitude of the applied voltage waveform and $F_{tri}(t)$ is the normalized waveform function of the driving signal (Figure 1b). The drift term is suppressed because the modulation rate far exceeds the drift relaxation rate $1/\tau_{drift}$.

The transmitted power through a high-Q Lorentzian cavity as a function of the instantaneous detuning between the fixed laser wavelength and the swept resonance is,

$$P_{trans}(t) \propto \frac{1}{1+4(\frac{\lambda_{laser}-\lambda_{res}(t)}{\Delta\lambda_{opt}})^2} \tag{3}$$

where $\Delta\lambda_{opt}$ is the optical full-width at half-maximum (FWHM) linewidth of the cavity resonance. As $V(t)$ sweeps linearly through zero, $\lambda_{res}(t)$ passes through $\lambda_{laser}$, and the transmitted power traces out the Lorentzian lineshape of the cavity in the time domain, generating a detectable electrical spectrum on the photodetector (Figure 1c).

The instantaneous sweep rate of the resonance wavelength is,

$$\frac{d\lambda_{res}}{dt}=\eta \cdot \frac{dV}{dt} \tag{4}$$

As shown in Figure 1d, the time-domain electrical spectrum $P_{trans}(t)$ therefore exhibits a Lorentzian temporal envelope, with a temporal $\Delta t_{FWHM}$ determined by the condition that the resonance detuning equals $\pm\Delta\lambda_{opt}/2$,

$$\Delta t_{FWHM}=\frac{\Delta\lambda_{opt}}{\eta \cdot \frac{dV}{dt}} \tag{5}$$

This expression establishes the direct correspondence between the temporal width of the oscilloscope-detected electrical spectrum and the fundamental cavity and EO parameters (Figure 1a).

Since the voltage sweeps linearly during each half-cycle of the triangular wave, there exists a one-to-one, linear mapping between the time axis and the voltage axis. The temporal $\Delta t_{FWHM}$ of the electrical spectrum corresponds directly to a voltage-domain $\Delta V_{FWHM}$ through,

$$\Delta V_{FWHM} = \frac{dV}{dt} \cdot \Delta t_{FWHM} \tag{6}$$

Substituting Eq. (5) into Eq. (6) yields,

$$\eta = \frac{\Delta\lambda_{opt}}{\Delta V_{FWHM}} \tag{7}$$

This equation enables direct extraction of the intrinsic EO tuning efficiency η using only two measurable quantities: (i) the static optical linewidth $\Delta\lambda_{opt}$, obtained from a Lorentzian fit to the cavity transmission spectrum prior to electrical driving, and (ii) the voltage-domain $\Delta V_{FWHM}$, converted from the temporal $\Delta t_{FWHM}$ of the oscilloscope-measured electrical spectrum via Eq. (6). On the linearly rising segment of the triangular wave, the voltage changes at a constant rate $dV/dt=2V_{pp}f$. Thus, the tuning efficiency η can be explicitly rewritten using the signal generator and time-domain parameters,

$$\eta = \frac{\Delta\lambda_{opt}}{2V_{pp} f \Delta t_{FWHM}} \tag{8}$$

Critically, this method eliminates the need for absolute wavelength tracking, and is inherently immune to drift-induced artifacts, as the quasi-static drift term does not alter the dynamic, short-timescale linewidth measurement.

## ■ DEVICE DESIGN AND FABRICATION

To validate this drift-free EO characterization methodology, a PCNBC modulator was designed and implemented on a polymer-loaded LNOI platform. The cavity structure follows our previously demonstrated high-Q PCNBC design27, which induces effective refractive index modulation via a patterned polymer layer atop the unetched LN film. The optical mode is strongly confined within the single-crystal LN core layer, while its evanescent field extends into the overlying polymer photonic crystal structure, facilitating strong and efficient EO interaction. The primary engineering focus of this work is the electrode design, as the electrode geometry directly governs the EO tuning efficiency η and must be optimized to maximize the overlap between the applied electric field and the confined optical mode. As schematically illustrated in Figure 2a, the electrodes feature a butterfly-shaped coplanar configuration arranged symmetrically about the central axis of the polymer waveguide. This design deliberately excludes redundant routing or interconnect lines to minimize parasitic capacitance and inductance. As depicted in the cross-sectional view in Figure 2c, the LN film has a thickness $t_{LN}$=300 nm, bonded on a 2 μm $SiO_2$ buried oxide layer, and the polymer loading layer (AR-P 6200) has a thickness $t_P$=400 nm. A y-cut crystal orientation was chosen so that the in-plane electric field applied by the coplanar electrodes is directed along the crystallographic z-axis. This configuration directly engages the largest EO tensor component of LN, thereby maximizing the overall EO tuning performance.

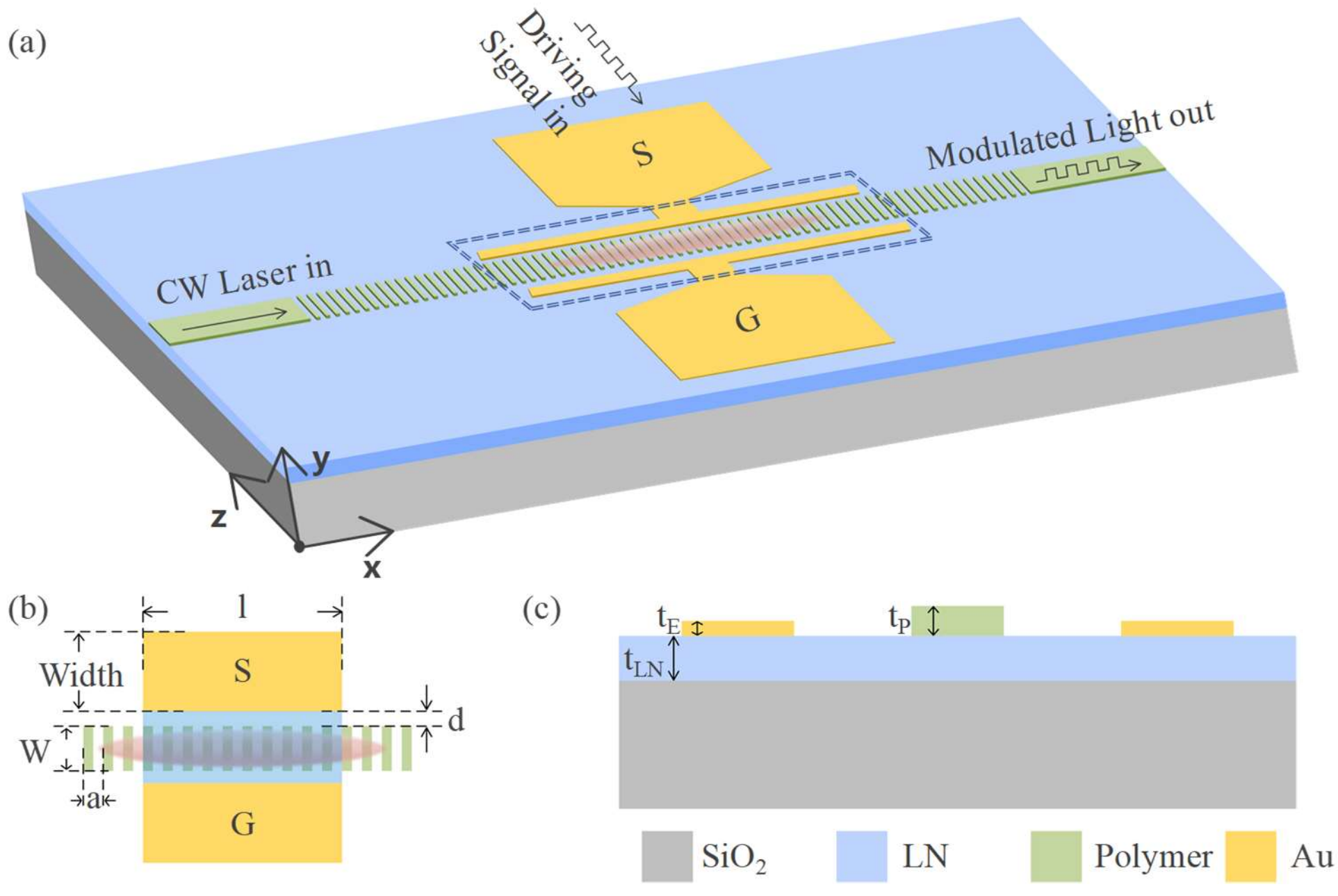

Figure 2. (a) Three-dimensional schematic of the integrated electro-optic modulator. (b) Two-dimensional top view of the active modulation region. The spatial overlap between the optical cavity mode (red profile) and the in-plane electric field (blue region) is determined by l (the electrode modulation length) and d (the electrode-to-waveguide lateral gap). (c) Cross-sectional schematic of the hybrid waveguide profile.

For a given cavity design, the EO tuning efficiency η is primarily governed by two distinct geometric electrode parameters. One is the modulation length (l), defined as the physical length of the electrode along the cavity axis, and the other is d, defined as the lateral separation between the inner edge of each electrode and the edge of the polymer waveguide. The physical roles of these two parameters are distinct and can be understood through the spatial overlap integral between the optical mode intensity distribution and the applied electric field.

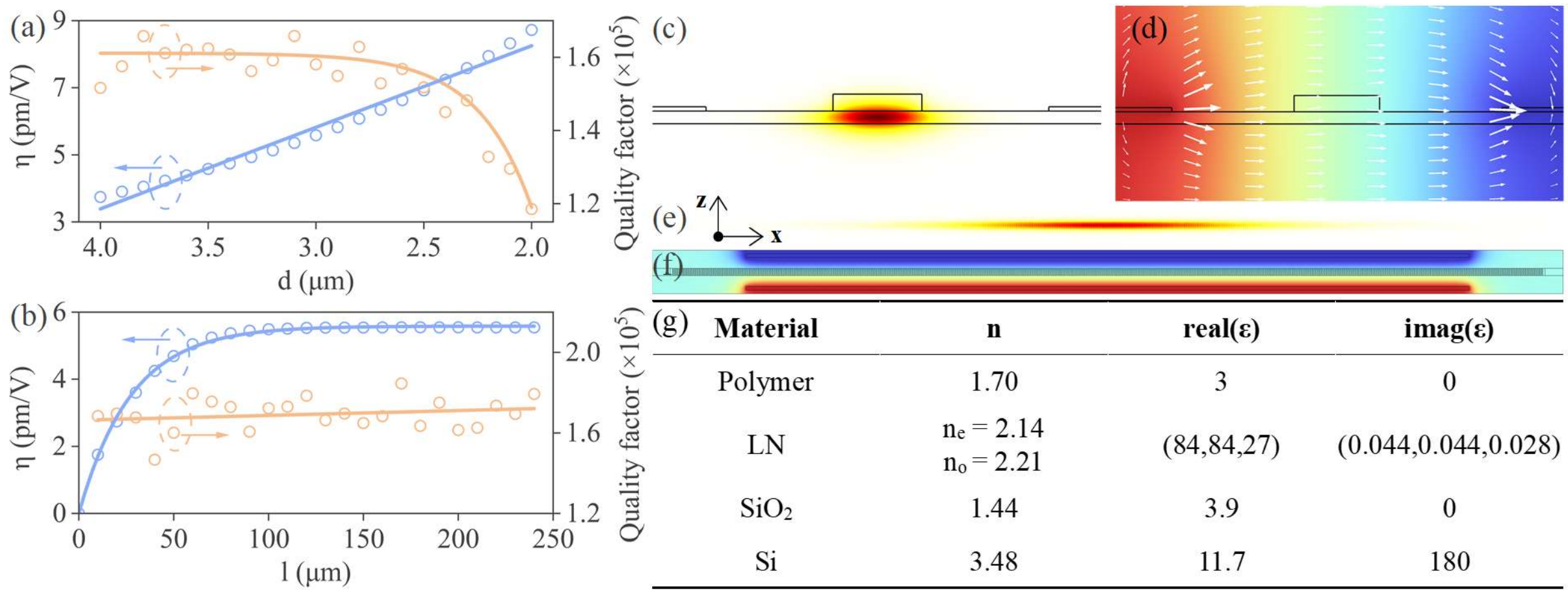

| Material | n | real(ε) | imag(ε) |
|---|---|---|---|
| Polymer | 1.70 | 3 | 0 |
| LN | $n_e = 2.14$<br>$n_o = 2.21$ | (84,84,27) | (0.044,0.044,0.028) |
| $SiO_2$ | 1.44 | 3.9 | 0 |
| Si | 3.48 | 11.7 | 180 |

Figure 3. (a) Calculated η and Q as a function of d, simulated at a fixed l of 200 μm. (b) Calculated η and Q as a function of l at a fixed d of 3 μm. (c) Two-dimensional TE optical field intensity distribution of the fundamental cavity mode. (d) Cross-sectional electrostatic potential distribution and corresponding electric field vectors (white arrows) within the thin-film LN core. (e) and (f) Top-view profiles of the optical mode and the electrostatic potential along the nanobeam propagation axis, respectively. (g) Summary table of the anisotropic refractive indices and complex relative permittivities used for each material layer in the 3D-FEM model.

To guide device optimization and provide quantitative theoretical benchmarks across the entire parameter space explored experimentally, comprehensive 2D and 3D finite-element method (FEM) simulations were performed using COMSOL Multiphysics. The simulations fully couple the electrostatic solver with the wave optics module. For each unique combination of l and electrode d, the spatial profiles of the applied electric field within the LN layer were evaluated. The local refractive index perturbation was calculated using the linear Pockels relation ($\Delta n_e = -(1/2) n_e^3 r_{33} E_z$).

The simulated η values and Q-factor for the first-order cavity modes are shown in Figure 3. The simulation results reveal two key trends both in EO tuning efficiency η and Q-factor in Figure 3a and b. In Figure 3a, for a fixed l with 200 μm, the tuning efficiency η increases approximately linearly with decreasing d (blue points). This is expected, as a larger d reduces the electric field strength within the LN layer for a given applied voltage. While a smaller d improves tuning efficiency η, it also introduces metal-induced optical loss due to field penetration into the electrodes. The results of Q-factor indicate that significant optical absorption occurs when $d \leq 3$ μm, whereas for $d \geq 3$ μm, metal-induced loss becomes negligible (orange points). Conversely, Figure 3b shows the scaling behavior with respect to l at a fixed d of 3 μm. The tuning efficiency η increases monotonically with l reaching a distinct saturation plateau at approximately 120 μm. Beyond this critical length, further extension of the electrodes yields no noticeable enhancement in η, demonstrating that l of 120 μm completely encompasses the localized optical mode volume of the fundamental PCNBC resonance.

Guided by these numerical results, an experimental matrix of 35 distinct devices was designed. The array systematically spans l of 10, 20, 40, 80, 120, 160, and 200 μm, combined with d of 3.1, 3.3, 3.5, 3.7, and 3.9 μm. This comprehensive device matrix allows for a rigorous quantitative comparison between the FEM-predicted values and the experimentally extracted tuning efficiencies obtained via the dynamic triangular-wave scanning method proposed in Section II, thereby offering a complete cross-validation loop for both the simulation framework and the characterization methodology.

All devices were fabricated on customized y-cut LNOI wafers supplied by NanoLN Corporation, comprising a 300 nm single-crystal $LiNbO_3$ device layer on a 2 μm thermally grown $SiO_2$ layer atop a silicon handle wafer. The fabrication flow proceeds in two sequential electron-beam lithography (EBL) steps with no etching of the LN film at any stage. The electrode pattern was defined in the first EBL step using a positive resist. After development, a bilayer metal stack consisting of 5 nm Ti and 95 nm Au was deposited by electron-beam evaporation (EBE) at a base pressure below 5×10-7 Torr to ensure film continuity and low sheet resistance. The lift-off process was performed by immersing the chip in N-Methyl-2-pyrrolidone (NMP) at 60°C for 30 minutes, followed by gentle agitation and rinse in isopropanol and deionized water, yielding clean electrode edges with no residual metal bridging across the gap. The butterfly-shaped electrode geometry, with the two arms symmetrically flanking the future cavity position, was verified by optical microscopy after lift-off process. The electrode width is 3 μm, and the fabricated d values span 3.1 to 3.9 μm in increments of 0.2 μm.

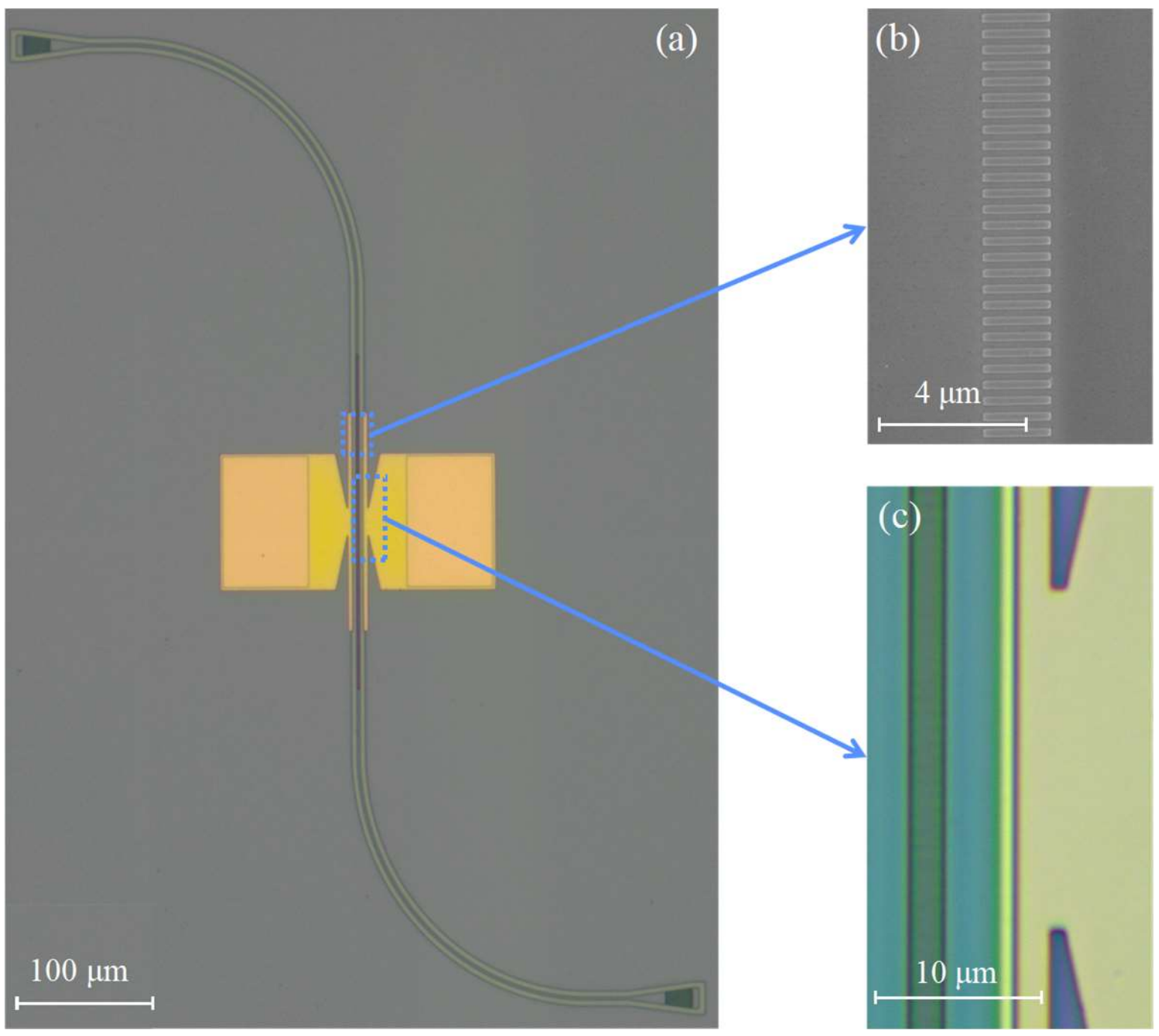

Figure 4. (a) Wide-field optical micrograph of the fabricated thin-film lithium niobate chip. (b) High-magnification SEM detailing the uniform periodic hole array and central defect region forming the polymer-loaded nanobeam cavity. (c) Zoomed-in optical micrograph focusing on the active modulation section.

A 400 nm thick film of AR-P 6200, a high-resolution positive-tone electron-beam resist selected for its optical transparency in the telecom band, its refractive index contrast with the LN surface, and its compatibility with the electrode metal surfaces already present on the wafer, was spin-coated at 4000 rpm and soft-baked at 150°C for 2 minutes. The second EBL exposure defined all optical components simultaneously: the fiber-to-chip grating couplers, the single-mode access waveguides, and the PCNBC with its taper and mirror unit cell arrays. The exposure dose was carefully calibrated to achieve the target filling factor profile. After development and rinsing, the patterned AR-P 6200 layer was retained as the permanent polymer loading material, forming the photonic crystal structure without any subsequent etching of the underlying LN film. Figure 4 displays optical and scanning electron micrographs of the completed devices, verifying the highly uniform definition of both the photonic crystal lattice and the electrodes.

## ■ RESULTS AND DISCUSSION

The experimental configuration used for EO characterization is outlined in Figure 5. A tunable semiconductor laser (TSL, Santec TSL-550) featuring a wavelength coverage across the 1500-1630 nm band and an intrinsic linewidth of <100 kHz serves as the narrow-linewidth optical source. This laser linewidth is more than an order of magnitude narrower than the typical resonance linewidth of the cavities under test, ensuring that laser phase noise does not introduce artificial broadening to the measured lineshapes. A fiber polarization controller aligns the input light to the fundamental transverse-electric (TE)-like mode of the waveguide to maximize the electro-optic overlap with the lateral electric field. Light is coupled into and out of the chip via fully etched grating couplers, yielding a single-grating efficiency of approximately 15%-20%. To mitigate thermo-optic resonance drifts induced by ambient temperature fluctuations, the photonic chip is mounted on a high-precision thermoelectric cooler (TEC) stage, which is actively stabilized at 27.0 ± 0.01°C throughout all measurements.

For dynamic characterization, a dual-channel arbitrary waveform generator (FY3900-10MHz) is employed to generate periodic triangular voltage waveforms. The RF signal from Channel 1 of the generator is delivered to the on-chip lumped-element electrodes via a high-bandwidth microwave probe, as shown in the optical micrograph inset of Figure 5. For this work, we employ triangular waves with frequencies in the range 10 kHz-1 MHz and peak-to-peak amplitudes of 20-60 V, selected to produce a total resonance shift of 1-2 cavity linewidths, ensuring complete sampling of the Lorentzian lineshape without inducing nonlinear optical effects.

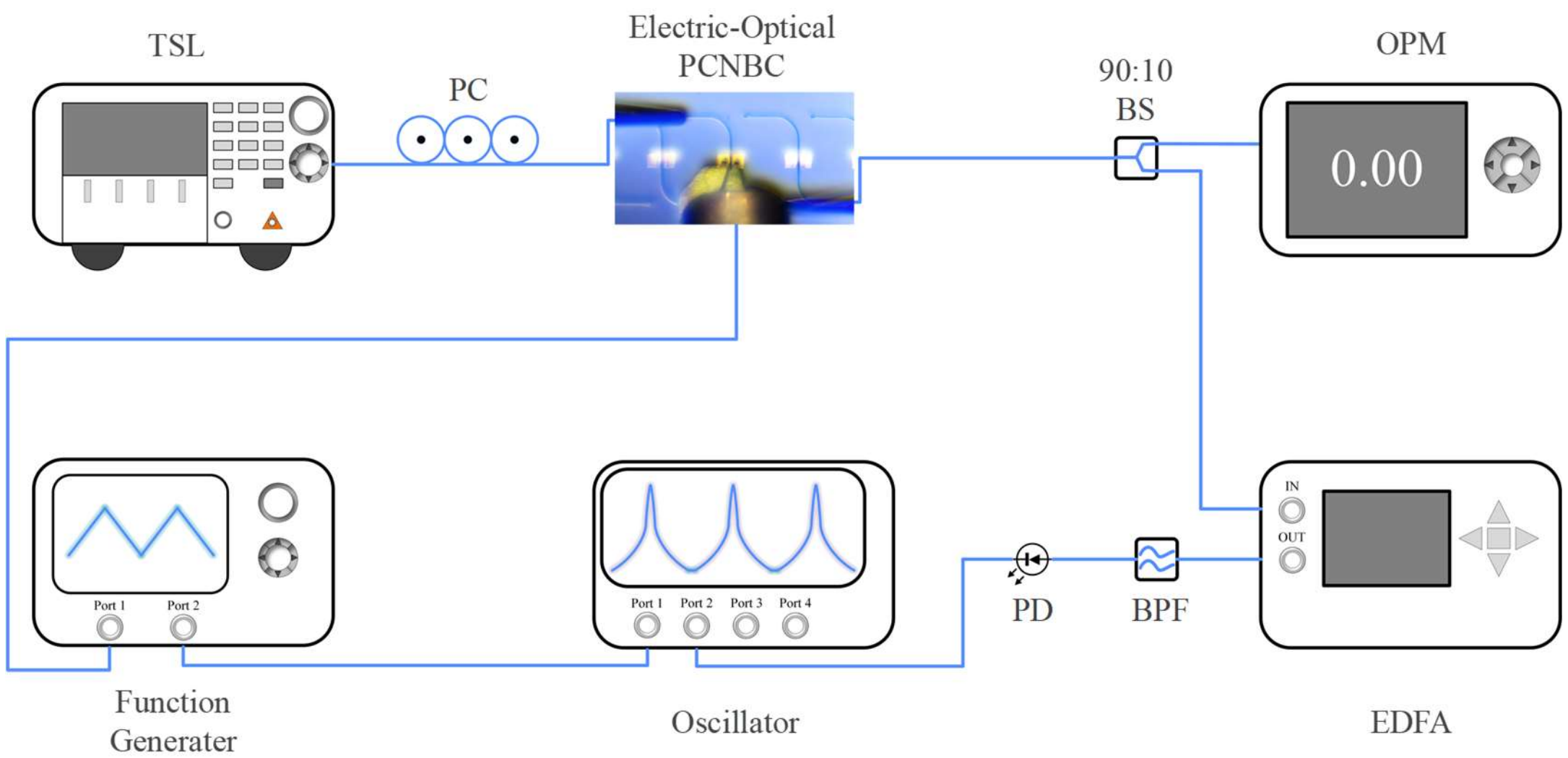

Figure 5. Schematic illustration of the experimental testing setup.

The modulated optical signal coupled out from the PCNBC is sent to a 90:10 fiber beam splitter. 10% of the split optical power is directed to an optical power meter (OPM) for real-time monitoring of the fiber-to-chip coupling state and in-situ adjustment of the laser wavelength to align with the cavity resonance. The remaining 90% of the optical power is amplified by an erbium-doped fiber amplifier (EDFA) to boost the signal level, ensuring high-sensitivity detection by the photodetector. To eliminate amplified spontaneous emission (ASE) noise and spurious interference pulses generated by the EDFA, a fixed band-pass filter (BPF) with a 30 nm bandwidth centered at 1560.00 nm is inserted between the EDFA and the photodetector; this bandwidth is sufficiently broad to fully cover the resonance wavelengths of all fabricated PCNBC devices while suppressing out-of-band ASE noise.

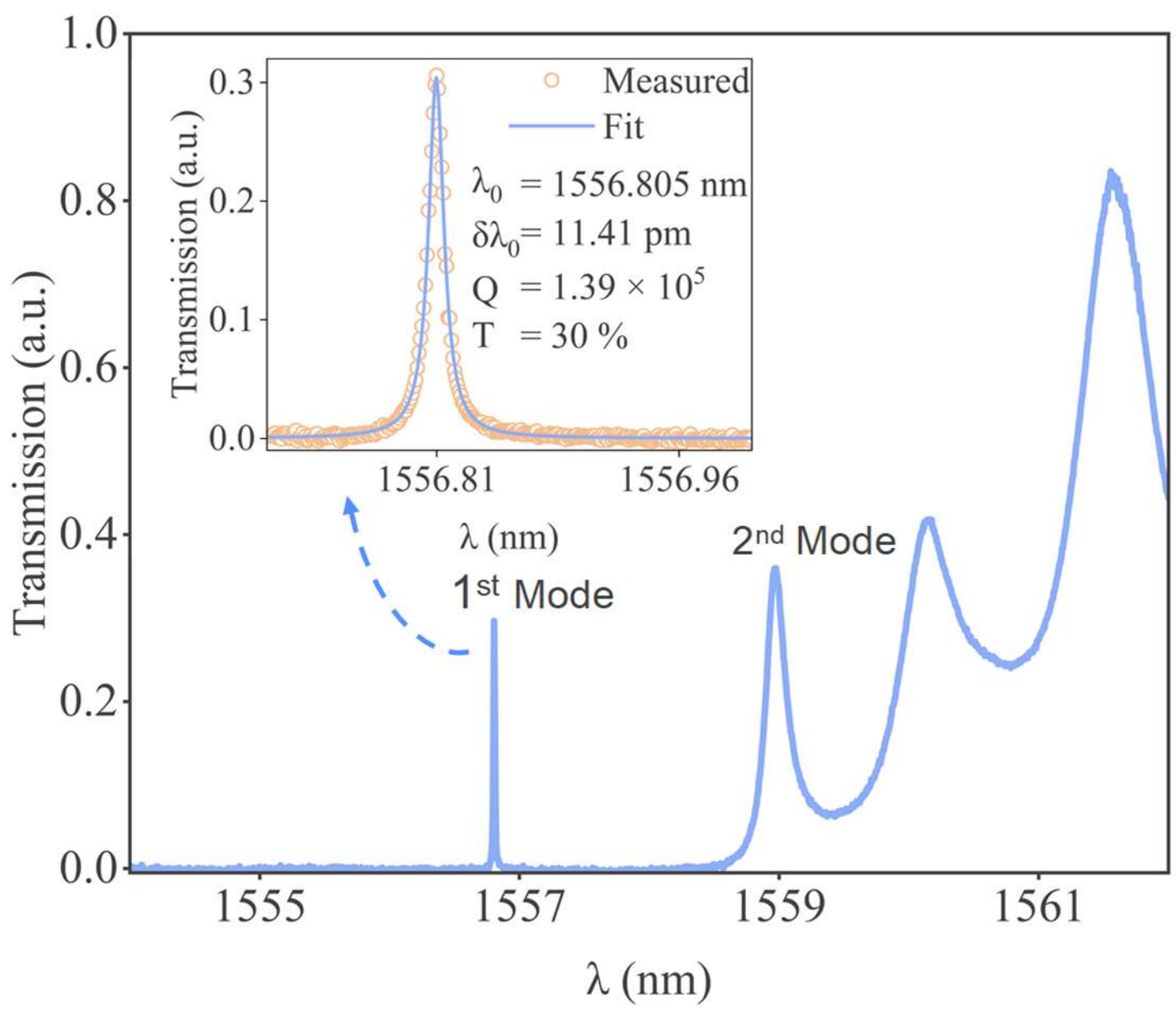

Figure 6. Measured transmission spectrum of the PCNBC. In the inset, the orange circles represent the enlarge view of the measured spectrum and the blue line represents the Lorentz fitting of the first-order mode.

The filtered optical signal is then detected by a high-speed InGaAs photodetector (PD, Nortel Networks PP-10G, 3 dB bandwidth > 10 GHz), whose electrical output is fed to Channel 1 of a high-bandwidth digital oscilloscope (Tektronix MSO64B, 4 GHz analog bandwidth, 50 GS/s sampling rate). To enable accurate determination of the actual voltage amplitude loaded onto the modulator electrodes, the synchronous RF signal from Channel 2 of the function generator (identical in waveform and amplitude to the driving signal from Channel 1) is connected to Channel 2 of the oscilloscope. Critically, Channel 2 of the oscilloscope is configured to a high-impedance input mode. Because the lumped-element design of the electro-optic modulator electrodes presents a high-impedance load at the kHz to low-MHz modulation frequencies employed in this work, the high-impedance oscilloscope input ensures accurate measurement of the true voltage amplitude applied across the device electrodes, eliminating errors from impedance mismatch. The oscilloscope is triggered synchronously via the function generator to ensure stable, phase-locked waveform capture of both the modulated optical signal and the reference driving voltage. Typical acquisition parameters include 1000-10000 waveform averages to reduce thermal and shot noise, yielding a final signal-to-noise ratio (SNR)>40 dB for all measurements.

For initial static optical characterization, the transmission spectrum is acquired by sweeping the TSL wavelength across the cavity resonance while monitoring the transmitted power on the OPM, with the RF drive disabled. The loaded Q-factor, extinction ratio, and resonance wavelength are extracted from a Lorentzian fit to the measured transmission spectrum. Figure 6 shows the transmission spectrum of a representative device, revealing multiple optical resonances corresponding to different transverse mode families. The fundamental mode resonance is centered at 1556.805 nm with an extinction ratio of 30% and a loaded Q-factor of 1.39×105, corresponding to an optical linewidth Δλopt of approximately 11.41 pm. This high Q-factor confirms the strong light confinement achieved by the polymer-loaded photonic crystal structure and provides the narrow linewidth necessary for sensitive electro-optic measurements. Once the resonance is identified, the laser wavelength is locked to the peak transmission point of the cavity for subsequent dynamic EO tuning efficiency η measurements.

To quantitatively demonstrate the fundamental unreliability of conventional DC scanning and validate the drift-immunity of the proposed triangular-wave methodology, we performed a systematic comparison across eight experimental groups for each characterization method on the same device (d = 3.1 μm, l = 120 μm). In each measurement set, a sequence of discrete bias voltages (+20,+10,0,-10,-20 V) is applied, while the interval between successive scans is systematically varied from 0 s to 120 s. As shown in Figures 7a and 7d, the extracted tuning efficiency η exhibits severe, irregular fluctuations as a function of the wait time, yielding a high coefficient of variation (CV) of 16.13%. This severe instability illustrates why quasi-static measurements provide unreliable assessments of resonant devices. In contrast, the proposed triangular-wave scanning method delivers exceptional stability and reproducibility. For this validation, the experimental variable across eight distinct test groups was the driving frequency of the triangular wave, spanning from 10 kHz to 1 MHz. Figure 7c displays the oscilloscope-detected time-domain transmission spectra under different modulation frequencies, and Figure 7b highlights five representative electrical spectra selected from each group for data processing. This frequency range is deliberately chosen to far exceed the characteristic drift frequency (typically $< 1$ kHz), ensuring that the instantaneous Pockels response can be cleanly decoupled from the quasi-static drift term, as discussed in Eq. (1) and Eq. (2). Across all test frequencies, the extracted tuning efficiency $\eta$ remains remarkably consistent within a tight window of 4.3-4.5 pm/V, yielding a highly reproducible average value of 4.39 pm/V with a remarkably low CV of 1.1% (Figure 7e). The exceptional consistency across a wide frequency range confirms that the dynamic sweep method effectively circumvents slow drift-induced artifacts, offering a robust and reliable pathway for extracting intrinsic EO parameters.

Having established the robustness of the drift-free methodology, we conducted a systematic study across a statistically significant population of 35 fabricated modulators to evaluate design-dependent scaling behaviors. The experimentally extracted tuning efficiencies as functions of $l$ and $d$ are plotted in Figures 8a and 8b, respectively. For a fixed $d$, the tuning efficiency $\eta$ increases monotonically with $l$ from 10 μm to approximately 120 μm, beyond which it gradually saturates. This behavior indicates that the effective interaction region between the optical mode and the applied electric field is fully covered when $l$ reaches ~120 μm. Further extension of the electrode length does not contribute additional overlap, confirming the spatial confinement of the fundamental cavity mode. For a fixed $l$, the tuning efficiency $\eta$ increases approximately linearly with decreasing $d$. This trend can be attributed to the increase in electric field intensity within the LN layer as the electrode separation reduces, leading to a stronger EO modulation effect. Importantly, these trends are consistently observed across the full device set, demonstrating strong reproducibility. Although a small number of outliers (four devices) are identified, these deviations are attributed to fabrication tolerances

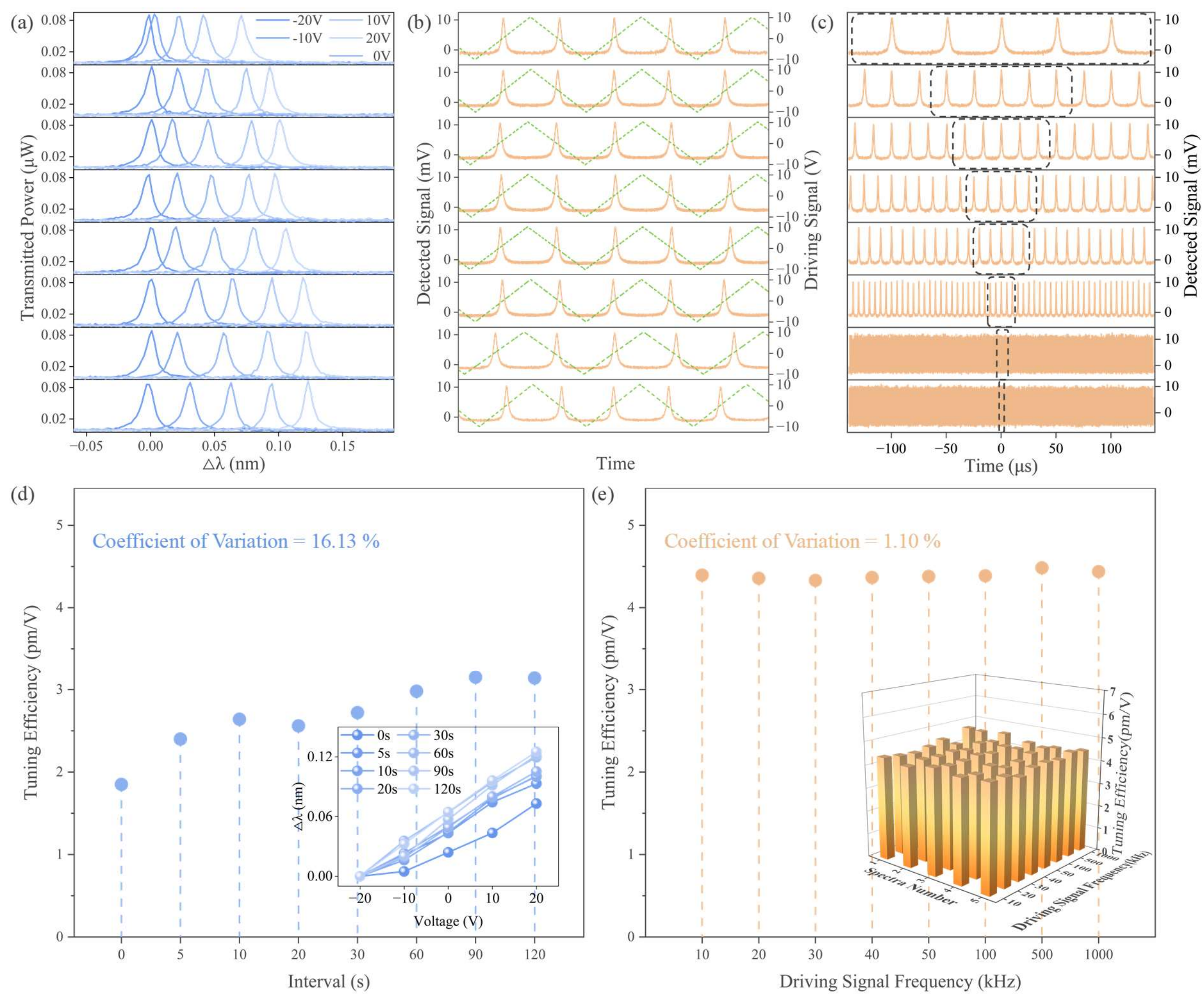

Figure 7. (a) Extracted resonance wavelength shifts under conventional quasi-static DC voltage scanning, plotted for wait intervals of 0 s and 120 s. (b) Five representative time-domain electrical spectra extracted from the dynamic sweeps for signal processing. (c) Raw time-domain transmission waveforms captured across eight distinct test groups by varying the driving triangular-wave frequency from 10 kHz to 1 MHz. (d) Measured tuning efficiency η derived via the conventional DC scanning method as a function of interval wait time. Inset: Tracking trajectory of the five resonance centers. (e) Calibrated tuning efficiency η extracted via the proposed dynamic triangular-wave sweep method as a function of modulation frequency. Inset: Highly uniform extracted η.

inherent to nanoscale lithography and do not affect the overall scaling behavior.

To validate the physical consistency of our results, the experimentally extracted tuning efficiencies were benchmarked against the 3D-FEM simulations described in Section III. The measured average tuning efficiency η exhibit excellent agreement with the numerical trends in both absolute magnitude and geometric scaling curves. On average, the simulated tuning values are slightly higher than the experimental data by a factor of approximately 0.2. This minor, systematic discrepancy is acceptable and common in integrated EO modulators. It arises because the idealized numerical model assumes perfectly smooth vertical walls and neglects secondary real-world factors such as slight processing-induced deviations in electrode positioning, variations in the spin-coated polymer thickness, and subtle sidewall roughness. Additionally, the ideal simulation model did not fully incorporate complex conductive and interfacial ohmic losses within the thin metal buffer layers, which can slightly attenuate the effective voltage dropped across the optical waveguide core in experimental settings.

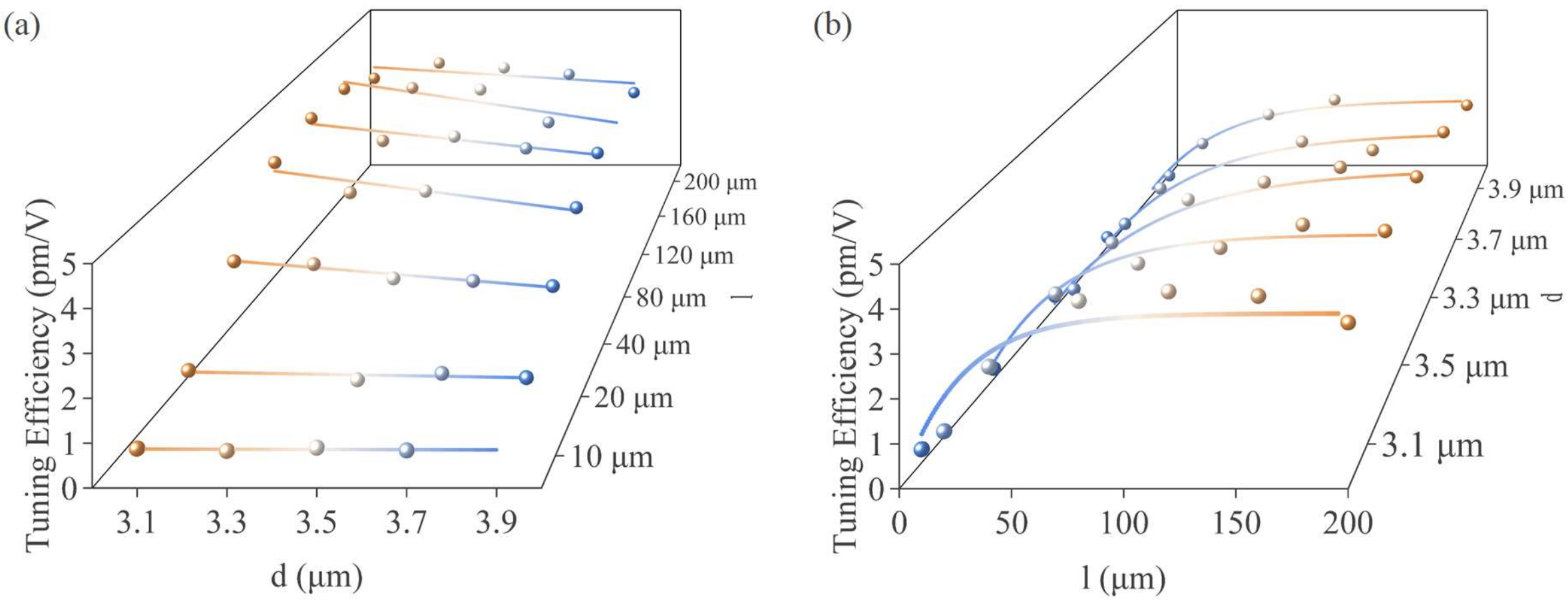

Figure 8. (a) Measured η plotted against the lateral d. (b) Measured η as a function of l across varying device configurations, demonstrating close agreement with the 3D-FEM simulated scaling profile and confirming that l of 120 μm fully encompasses the fundamental cavity mode volume.

Beyond this specific platform, the demonstrated methodology provides a general and scalable framework for drift-free EO characterization in high-Q resonant photonic devices. By eliminating the need for absolute wavelength tracking and relying instead on time-domain linewidth extraction, this approach is inherently robust against slow environmental and material instabilities. It is therefore readily extendable to other cavity-based modulators, where DC drift similarly limits measurement accuracy.

## ■ CONCLUSION

In conclusion, we have systematically investigated the DC drift effect in polymer-loaded LN PCNBC EO modulators and demonstrated a robust, drift-free characterization framework using high-frequency triangular-wave voltage sweeps. This approach exploits the distinct separation in relaxation timescales between the instantaneous Pockels effect and slow interfacial charge dynamics, achieving highly reproducible tuning efficiency η extractions with a coefficient of variation of 1.1%. Experimental validation across a statistical sample of 35 devices featuring diverse electrode geometries shows strong quantitative alignment with three-dimensional finite-element simulations. By eliminating the necessity for absolute resonance wavelength tracking, this technique offers a precise, standardized tool for evaluating cavity-based EO devices, thereby advancing their integration into next-generation photonic integrated circuits.

**Authors Contributions**

The manuscript was written through contributions of all authors. All authors have given approval to the final version of the manuscript.

**Notes**

The authors declare no competing financial interest.

**ACKNOWLEDGMENT**

The authors acknowledge support from the Natural Science Basic Research Program of Shaanxi (Program No. 2025JC-YBMS-666), CIE-Smartchip Research Fund (No. 2024-11), and the Fundamental Research Funds for the Central Universities (No. YJSJ26021).